\documentclass[twocolumn,prl,showpacs,superscriptaddress,showpacs]{revtex4-1}
\usepackage{graphicx}
\usepackage{amssymb,amsmath}
\usepackage{bm}
\usepackage{dcolumn}
\usepackage{subfigure}
\usepackage{float}
\usepackage{url}
\usepackage{color}
\usepackage{txfonts}
\usepackage[driverfallback=dvipdfm]{hyperref}
\hypersetup{pdfpagemode=FullScreen,colorlinks=true,breaklinks,urlcolor=blue,linkcolor=blue,citecolor=blue}
\usepackage{natbib}

\begin{document}

\title{Hall Conductance of a Non-Hermitian Chern Insulator}

\author{Yu Chen}
\affiliation{Center for Theoretical Physics, Department of Physics, Capital Normal University, Beijing 100048, China}
\author{Hui Zhai}
\affiliation{Institute for Advanced Study, Tsinghua University, Beijing, 100084, China}
\affiliation{Collaborative Innovation Center of Quantum Matter, Beijing, 100084, China}

\date{\today }

\begin{abstract}

In this letter we study the Hall conductance for a non-Hermitian Chern insulator and quantitatively describe how the Hall conductance deviates from a quantized value. We show the effects of the non-Hermitian terms on the Hall conductance are two folds. On one hand, it broadens the density-of-state of each band, because of which there always exists a non-universal bulk contribution. On the other hand, it adds decay term to the edge state, because of which the topological contribution also deviates from the quantized Chern number. We provides a simple formula for the topological contribution for a general two-band non-Hermitian Chern insulator, as a non-Hermitian version of the Thouless-Kohmoto-Nightingale-de Nijs formula. It shows that the derivation from quantized value increases either when the strength of the non-Hermitian term increases, or when the momentum dependence of the non-Hermitian term increases. Our results can be directly verified in synthetic non-Hermitian topological systems where the strength of the non-Hermitian terms can be controlled.  
 
\end{abstract}

\maketitle

In the past decades, topological band theory has been extensively studied not only in electronic system \cite{Kane10,Zhang11,Das16,Ryu16,Vishwanath18}, but also in synthetic cold atom, photonic and acoustics systems \cite{Zoller16,Spielman18,rev_photon}. Nevertheless, an aspect that has only been brought out recently is the interplay between non-Hermitian and topology \cite{Shen,SSHa,SSHb,SSHc,SSHd,SSHe,Wang,Ueda,Levitov09,Hughes11,Kohmoto11,Huang13,Schomerus13,Schomerus15,Aguado16,Lee16a,Duan17,Menke17,Molina17,Haas17,Lieu18,Zyuzin18,Fan18,Torres18,Molina18,FuArc17,Ueda18a,Khanikaev18,Fu18,Ueda18c,Yoshida18,Szameit15,Szameit17,Obuse17,Fu18ex,Feng18,Parto18,TopLaserT,TopLaserE,Feng18b}. In fact, the non-Hermitian nature exists generically in all realization of topological band theory mentioned above. The non-Hermitianess can come from the imaginary part of the self-energy and finite life time of quasi-particles in electronic systems, loss of atoms in cold atom realization, optical gain and loss of photons in photonic realization and the damping of mechanical modes in acoustics systems.

In a Hermitian system, the most profound manifestation of the topological band theory is the quantum Hall effect, which states that the Hall conductance for a gapped band insulator is quantized and equals to the sum of the Chern number of all filled bands. This is expressed as the famous Thouless-Kohmoto-Nightingale-de Nijs (TKNN) formula \cite{TKNN}. It can also been shown that the number of gapless edge modes also equals to the net Chern number of all filled band, and because the absence of dissipation for transport through the edge state, the quantized Hall conductance is attributed to the conducting of electrons through these quantized edge modes when bulk is gapped. 

So far most study of the non-Hermitian topological band theory focus on solving the Schr\"odinger equation and obtain properties from the eigen-energies and the eigen-vector of a non-Hermitian Hamiltonian. For instance, the generalization of the ``gapped bands" to ``separable bands" is based on the distribution of the eigen-energies on the complex plane \cite{Shen}. The topological invariant defined from eigen-energies that is unique to separable non-Hermitian Hamiltonians is found to protect bulk Fermi arcs \cite{Shen,FuArc17,Fu18ex}, and solving the topologically nontrivial non-Hermitian Hamiltonian with the open boundary condition can lead to boundary modes for separable bands \cite{Shen, Ueda}. More interestingly, it is found in some non-Hermitian Hamiltonians the bulk spectrum is extremely sensitive to the boundary condition so that the conventional bulk-edge correspondence can breakdown \cite{SSHa,SSHb,SSHc,SSHd,SSHe,Wang,Ueda}. The classification of non-Hermitian topological Hamiltonian has also been discussed \cite{Ueda}. 

In this letter we concentrate on physical observables, in particular, the Hall conductance. We shall first clarify the meaning of ``gapped insulator" and we emphasize that a gapped insulator means the density-of-state should be sufficiently small at the chemical potential. Nevertheless, the residual density-of-state inside the ``gap" can always give rise to finite bulk contribution to the Hall conductance. Then we work out the formula for the topological contribution of the Hall conductance as a non-Hermitian generalization of TKNN formula, with which we can quantify how the Hall conductance deviates from a quantized value even the topological index defined for non-Hermitian Hamiltonian is a non-zero integer. In another word, unlike the Hermitian case, the existence of quantized edge state does not necessarily lead to a quantized Hall conductance in the non-Hermitian case. Even a nearly quantized Hall conductance requires more strict condition than a non-zero non-Hermitian Chern number, which we will specify as the main results of this work.  

\textit{Green's function.} Let us start with the general discussion of the Green's function for a non-Hermitian Hamiltonian $\hat{H}$. We assume $|\varphi^R_i\rangle$ and $\langle \varphi^L_i|$ are the right and the left eigen-vectors of $\hat{H}$ with the same complex eigen-value $\epsilon_i$, that is
\begin{eqnarray}
\hat{H}|\varphi^R_{i}\rangle=\epsilon_i|\varphi^R_{i}\rangle, \hspace{3ex}
\langle\varphi^L_i|\hat{H}=\epsilon_i\langle\varphi^L_i|.
\end{eqnarray}
It can be shown that, as long as $\epsilon_i\neq\epsilon_j$, we have $\langle\varphi^L_i|\varphi^R_{j}\rangle=0$. As long as $\langle\varphi^L_i|\varphi^R_{i}\rangle\neq0$, we can construct an operator $\hat{P}_i$ as
\begin{eqnarray}
\hat{P}_i\equiv\frac{1}{\langle\varphi^L_i|\varphi^R_{i}\rangle}|\varphi_i^R\rangle\langle\varphi_i^L|.
\end{eqnarray}
It can be proved that $\hat{P}_i^2=\hat{P}_i$ and $\hat{P}_i\hat{P}_j=0$ for all $ i\neq j$, which shows that $\hat{P}_i$ is in fact a projection operator. Furthermore, one can show that $\sum_i\hat{P}_i=\hat{I}$. The Hamiltonian can be expressed as $\hat{H}=\sum_{i}\epsilon_{i}\hat{P}_i$ when there is no degenerate eigen-energies. 

Here we consider the situation that the imaginary parts of all $\epsilon_i$ are negative, which is true for non-Hermitian models that are resulted from coupling the system to an environment \cite{Midtgaard}. For this case, with these conditions for $\hat{P}_i$, we can show that the Green's function can be written as 
\begin{eqnarray}
\hat{G}^\text{R}=(\epsilon-\hat{H})^{-1}=\sum_i\frac{\hat{P}_i}{\epsilon-\epsilon_i} \label{Lehmann}
\end{eqnarray}
In the Hermitian case, the Lehmann's representation for the Green's function takes the same form as Eq. \ref{Lehmann} except $\hat{P}_i$ there is defined as $|\varphi_i\rangle\langle \varphi_i|$, where $|\varphi_i\rangle$ is normalized eigen-states. Thus, Eq. \ref{Lehmann} be viewed as the non-Hermitian version of the Lehmann's representation of the Green's function. To prove this, one only needs to verify that $(\epsilon-\hat{H})\hat{G}^\text{R}=\hat{I}$ when $\hat{G}^\text{R}$ is expressed as Eq. \ref{Lehmann}, which can follow from straightforward derivation as 
\begin{eqnarray}
&&(\epsilon-\hat{H})\hat{G}^\text{R}=\sum_j(\epsilon-\epsilon_j)\hat{P}_j\sum_i\hat{P}_i\frac{1}{\epsilon-\epsilon_i}\nonumber\\
&&=\sum_{ij}\hat{P}_i\hat{P}_j(\epsilon-\epsilon_j)(\epsilon-\epsilon_i)^{-1}=\sum_i\hat{P}_i=\hat{I}
\end{eqnarray}
where we have used $\hat{P}_i\hat{P}_j=\delta_{ij}\hat{P}_i$ in the last step.

\begin{figure}[t]\label{Fig:Gap}
\includegraphics[width=7cm]{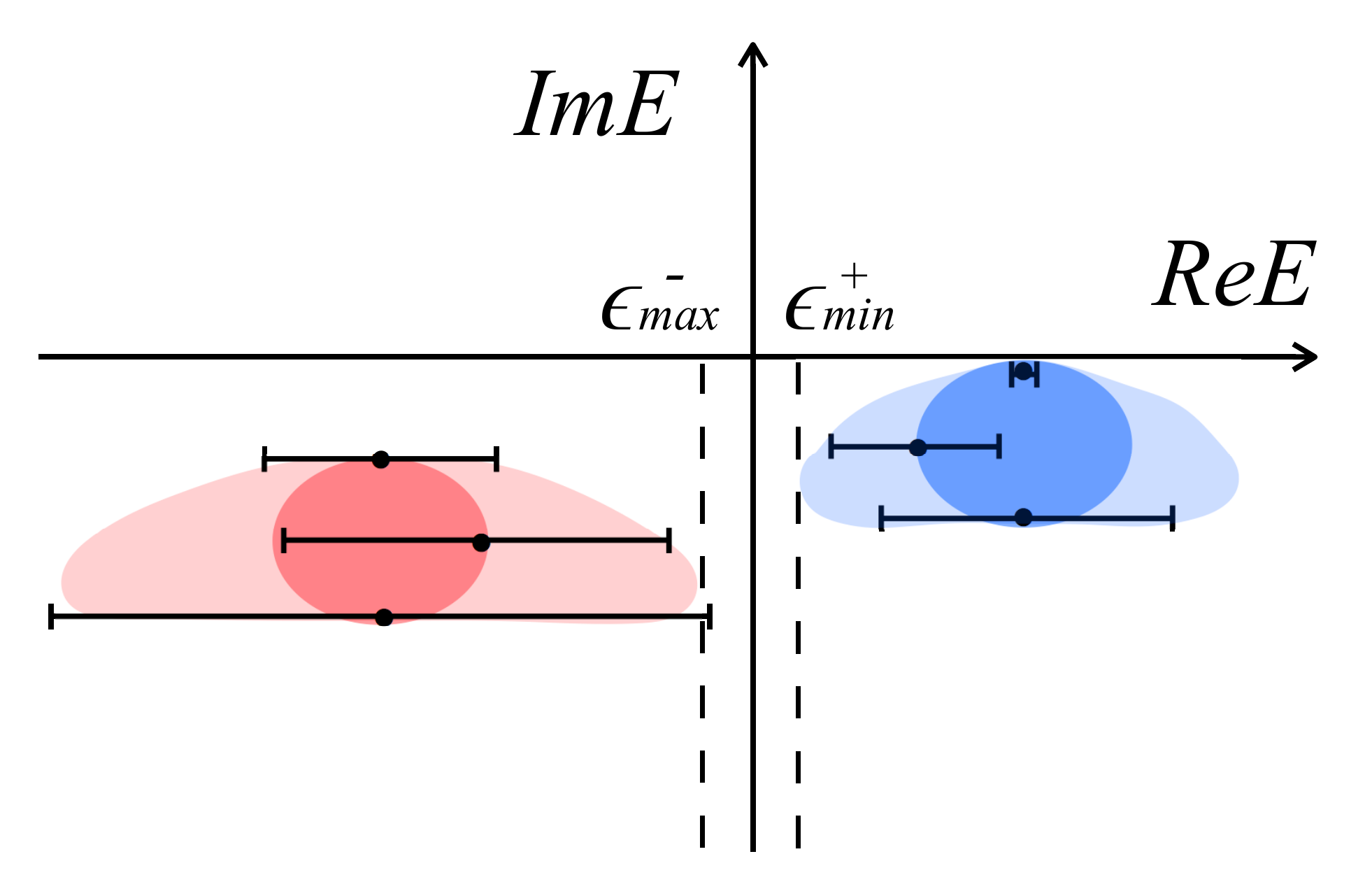}
\caption{An illustration of a gap. According to the spectrum decomposition theorem, each eigenstate with a complex energy $\epsilon_i$ contributes significant density-of-state from $[\epsilon_i^{\rm R}-|\epsilon_i^{\rm I}|,\epsilon_i^{\rm R}+|\epsilon_i^{\rm I}|]$ as is illustrated above. Therefore a gap could be defined as $\epsilon^+_\text{min}>\epsilon^-_\text{max}$ with $\epsilon^m_\text{max}$ and $\epsilon^m_\text{min}$ defined in Eqn. (6) and (7).  \label{gap}}
\end{figure}

Since the imaginary part of the retarded Green's function has the physical meaning as the density of states, therefore we have the density of state as $\rho(\epsilon)=-\pi A(\epsilon)$, and 
\begin{eqnarray}
A(\epsilon)&&={\rm Im}\text{Tr}\hat{G}^\text{R}=\sum_i{\rm Im}\frac{1}{\epsilon-\epsilon_i+i0^+}\nonumber\\
&&=\sum\limits_{i}\frac{\text{Im}\epsilon_i}{(\epsilon-\text{Re}\epsilon_i)^2+(\text{Im}\epsilon_i)^2}. \label{A}
\end{eqnarray}
That is to say, each $\epsilon_i$ contributes a Lorentzian shaped density-of-state centered at $\text{Re}\epsilon_i$ and a half-width $|\text{Im}\epsilon_i|$, as shown in Fig. \ref{gap}. Suppose all eigen-energies $\epsilon_i$ are separated into different bands labelled by $m$ and for each band the eigen-energies occupy a regime in the complex plane, even though there is no overlap between different regimes, it can not guarantee that the system is gapped. Let us now define for each band $m$, 
\begin{eqnarray}
\epsilon^m_\text{min}=\text{min}_{i\in m}(\text{Re}\epsilon_i-|\text{Im}\epsilon_i|),\\
\epsilon^m_\text{max}=\text{max}_{i\in m}(\text{Re}\epsilon_i+|\text{Im}\epsilon_i|),
\end{eqnarray} 
the contribution of band-$m$ to the density-of-state decays out only when $\epsilon\gg \epsilon^m_\text{max}$ or when $\epsilon\ll \epsilon^m_\text{min}$. Strictly speaking, there is no real gap because everywhere the density-of-state is finite. Nevertheless, when the ranges $[\epsilon^m_\text{min}, \epsilon^m_\text{max}]$ and $[\epsilon^{m^\prime}_\text{min}, \epsilon^{m^\prime}_\text{max}]$ are well separated, one can find a regime where the density-of-state contributed from all bands is sufficiently small. In another word, because each band is broadened by an energy scale of the imaginary part of the eigen-energies, in order for such a ``gap" to be defined, the minimum separation between the real part of the eigen-energies has to be much larger than the typical value of the imaginary part of the eigen-energies.
 
\textit{Hall Conductance.} With the expression for the retarded Green's function, we can proceed to discuss the Hall conductance with linear response theory. To be concrete, we focus on a two-band Chern insulator, whose Hamiltonian is written as
\begin{equation}
\hat{H}=\sum\limits_{{\bf k}}\hat{\Psi}_{{\bf k}}^\dag \mathcal{H}_{{\bf k}}\hat{\Psi}_{{\bf k}}
\end{equation}
where $\hat{\Psi}^\dag_{{\bf k}}=(\hat{c}^\dag_{{\bf k},\uparrow},\hat{c}^\dag_{{\bf k},\downarrow})$. $\mathcal{H}_{{\bf k}}$ is a $2\times 2$ matrix with a general form given by \cite{Hall18_const}
\begin{equation}
\mathcal{H}_{{\bf k}}=(d^0_{{\bf k}}-i\gamma^0_{{\bf k}})\times I +{\bf h}_{{\bf k}}\cdot \boldsymbol{\sigma}
\end{equation}
and ${\bf h}_{{\bf k}}={\bf d}_{{\bf k}}+i \boldsymbol{\gamma}_{{\bf k}}$, and both ${\bf d}_{{\bf k}}=(d^x_{{\bf k}},d^y_{{\bf k}},d^z_{{\bf k}})$ and $\boldsymbol{\gamma}_{{\bf k}}=(\gamma^x_{{\bf k}},\gamma^y_{{\bf k}},\gamma^z_{{\bf k}})$ are three component vectors. For each ${{\bf k}}$, the eigen-energies of this matrix is given by $\epsilon^{\pm}_{{\bf k}}=d^0_{{\bf k}}\pm \lambda_{{\bf k}}-i\gamma^0_{{\bf k}}$, where $\lambda_{{\bf k}}=\sqrt{\sum_{i=x,y,z} h_{{\bf k}}^{i,2}}$ is in general a complex number and we denote it as $\lambda_{{\bf k}}=\lambda^\text{R}_{{\bf k}}+i\lambda^{I}_{{\bf k}}$. Here we consider the situation that $\gamma^0_{{\bf k}}>0$ and $\gamma^0_{{\bf k}}$ is always greater than $|\lambda^{I}_{{\bf k}}|$ such that both $\epsilon^{\pm}_{{\bf k}}$ have negative imaginary parts. 

In this case, the retarded Green's function is expressed as 
\begin{equation}
\hat{G}^\text{R}=\sum\limits_{s=\pm,{\bf k}}\frac{\hat{P}^s_{{\bf k}}}{\epsilon-\epsilon^{s}_{{\bf k}}}, \label{GRk}
\end{equation} 
where $\hat{P}^{s=\pm}_{{\bf k}}=|u^\text{R}_{s,{\bf k}}\rangle\langle u^\text{L}_{s,{\bf k}}|$. Here we introduce $|\varphi^\text{R/L}_{s,{\bf k}}\rangle$ as the right/left eigenvector of $\mathcal{H}_{{\bf k}}$, and $|u^\text{R/L}_{s,{\bf k}}\rangle$ is normalized as 
\begin{equation}
|u^\text{R/L}_{s,{\bf k}}\rangle=\frac{|\varphi^\text{R/L}_{s,{\bf k}}\rangle}{\sqrt{\langle\varphi^L_i|\varphi^R_{i}\rangle}}.
\end{equation}
By noticing that $\mathcal{H}_{{\bf k}}$ can be written as
\begin{equation}
\mathcal{H}_{{\bf k}}=\epsilon_{{\bf k}}^{+}\frac{1+\hat{\bf {h}}_{{\bf k}}\cdot \boldsymbol{\sigma}}{2}+\epsilon_{{\bf k}}^{-}\frac{1-\hat{{\bf h}}_{{\bf k}}\cdot \boldsymbol{\sigma}}{2},
\end{equation}
where $ \hat{\bf {h}}_{{\bf k}}$ is defined as ${\bf h}_{{\bf k}}/\lambda_{{\bf k}}$,
we can also write
\begin{equation}
\hat{P}^s_{{\bf k}}=\frac{1}{2}(1\pm \hat{\bf {h}}_{{\bf k}}\cdot \boldsymbol{\sigma}). \label{PSk}
\end{equation} 

We define the current operator $\hat{J}_\alpha$ ($\alpha=x,y$) in this case as
\begin{equation}
\hat{J}_\alpha=\frac{\partial \text{Re}\mathcal{H}_{{\bf k}}}{\partial k_\alpha}=\partial_{k_\alpha}d^0_{{\bf k}}+(\partial_{k_\alpha}{\bf d}_{{\bf k}})\cdot  \boldsymbol{\sigma}. \label{Jk}
\end{equation}
Thus, we introduce the current-current correlation function 
\begin{equation}
K_{\alpha\beta}(\omega)=\sum\limits_{{\bf k}}\int d\epsilon d \epsilon^\prime\frac{n_\text{F}(\epsilon^\prime)-n_\text{F}(\epsilon)}{\epsilon^\prime-\epsilon+\omega+i0^+}\text{Tr}(\hat{J}_\alpha A(\epsilon)\hat{J}_\beta A(\epsilon^\prime)), \label{K}
\end{equation}
where $n_\text{F}(\epsilon)=1/(e^{\beta (\epsilon-\mu)}+1)$ ($\beta=1/(k_\text{B}T)$), 
and the conductance is defined as
\begin{equation}
\sigma_{\alpha\beta}=\lim\limits_{\omega\rightarrow 0}\frac{1}{\omega+i0^+}(K_{\alpha\beta}(\omega)-K_{\alpha\beta}(0)).
\end{equation}

With Eq. \ref{GRk}, Eq. \ref{PSk} and Eq. \ref{Jk}, both $\hat{J}$ and $A(\epsilon)$ used for calculating conductance with Eq. \ref{K} are explicitly known, and it is straightforward to calculate the Hall conductance and we ignore the tedious but straightforward intermediates steps here. Here we only focus on the case that these eigen-energies are separable into two regimes denoted by $\pm$ bands. 
Without loss of generality, we assume that $\text{Re}\epsilon^{+}_{{\bf k}}$ are all positive, and $\text{Re}\epsilon^{-}_{{\bf k}}$ are all negative such that both $\lambda^\text{R}_{{\bf k}}\pm h^0_{{\bf k}}$ are both positive. Thus, we can set the chemical potential $\mu=0$ at the band gap. Under these conditions, the final result for the zero-temperature Hall conductance $\sigma_\text{H}=\sigma_{xy}$ can be divided into two parts as $\sigma_\text{H}=\sigma_\text{H}^\text{T}+\sigma_\text{H}^\text{B}$, where $\sigma_\text{H}^\text{T}$ and $\sigma_\text{H}^\text{B}$ denote topological contribution and non-universal bulk contribution, respectively. $\sigma_\text{H}^\text{T}$ is given by  
\begin{equation}
\sigma_\text{H}^\text{T}=\sum\limits_{{\bf k}}\frac{\Omega_{xy}({\bf k})+\Omega^*_{xy}({\bf k})}{2}\times\nu_{{\bf k}} \label{Hall_cond}
\end{equation}
where $\Omega_{xy}({\bf k})={\bf h}_{{\bf k}}\times (\partial_{k_x}{\bf d}_{{\bf k}}\times \partial_{k_y}{\bf d}_{{\bf k}})/|{\bf h}_{{\bf k}}|^3$ and
\begin{equation}
\nu_{{\bf k}}=\frac{1}{\pi}\left(\text{arctan}\frac{\lambda^\text{R}_{{\bf k}}+h^0_{{\bf k}}}{|\gamma^0_{{\bf k}}-\lambda^\text{I}_{{\bf k}}|}
+\text{arctan}\frac{\lambda^\text{R}_{{\bf k}}-h^0_{{\bf k}}}{|\gamma^0_{{\bf k}}+\lambda^\text{I}_{{\bf k}}|}\right)
\end{equation}
This is the key result of this work.

\textit{Condition for a Nearly-Quantized Hall Conductance.} The formula for the Hall conductance Eq. \ref{Hall_cond} tells us that the Hall conductance for a non-Hermitian topological Chern insulator is in general not quantized. This by itself is not a surprise, and the reasons are two fold. On one hand, as discussed above, the bulk contribution to the density-of-state is always finite everywhere, which contributes a finite and non-universal Hall conductance denoted by $\sigma_\text{H}^\text{B}$. We do not include the detail form of these terms because they are quite complicated and depends on all details of the Hamiltonian, but it is included in the numerical calculation presented below. The general trend is that the smaller $\rho(\mu)$, the smaller this contribution. Thus, when two bands are sufficiently separated, as discussed above, the contribution from these terms can be strongly suppressed. On the other hand, because the energy for the edge state also possesses an imaginary part which means that the current can decay during the edge transport, because of which each edge state can only give a conductance less than one unit of quantized conductance. This effect is given by the term presented in Eq. \ref{Hall_cond}. The important aspect of our formula Eq. \ref{Hall_cond} is because it quantitatively characterizes how the contribution to the Hal conductance from the topological edge current deviates from the quantized value due to the non-Hermitian effect.   
\begin{figure}[t]
\includegraphics[width=8.2cm]{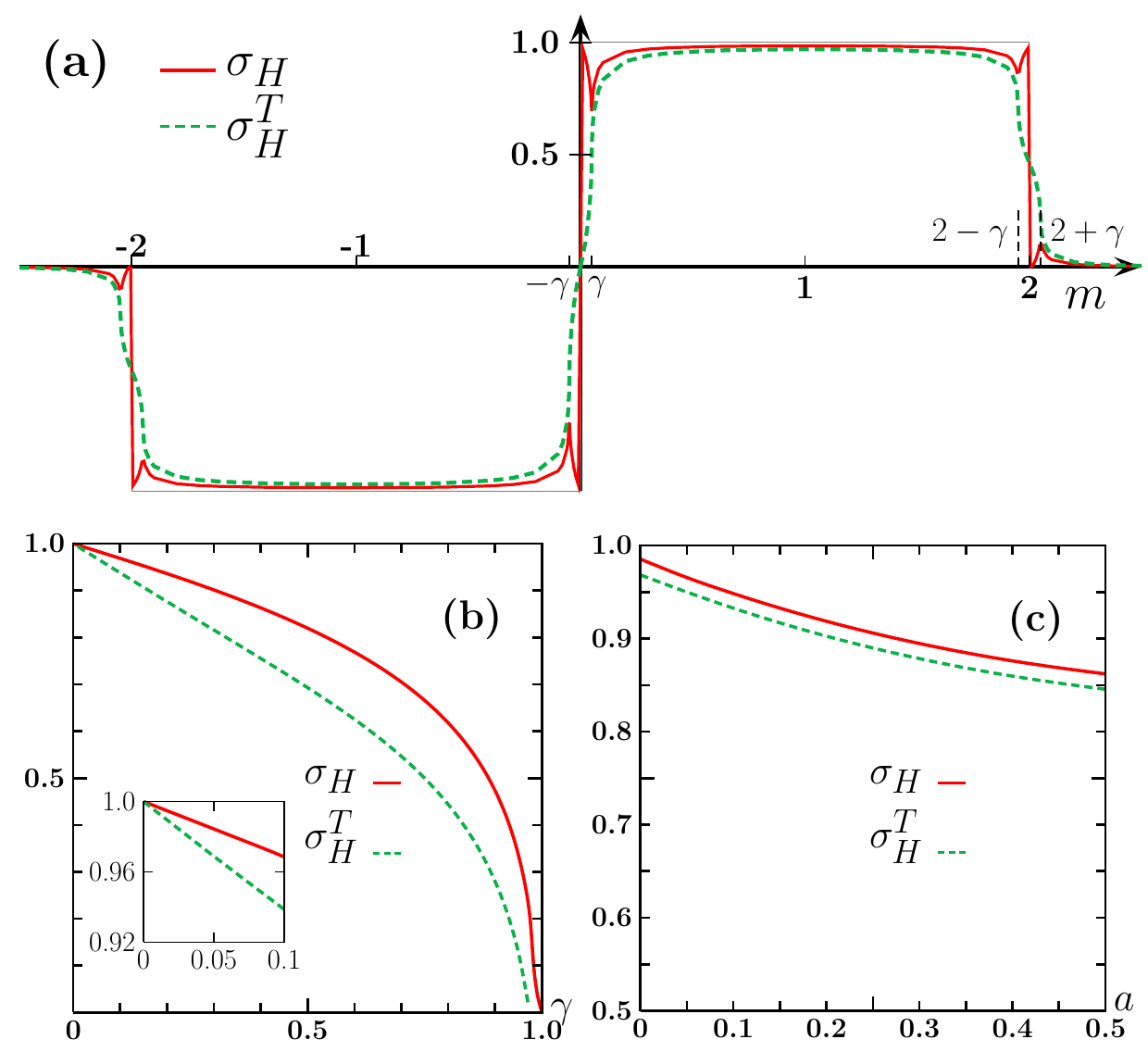}
\caption{The Hall conducance for the non-Hermitian Hamiltonian given by Eqn. \ref{Eqn:Model}. The total Hall conductivity $\sigma_H$ calculated numerically is given by the red solid lines and the topological contribution $\sigma_\text{H}^\text{T}$ given by Eq. \ref{Hall_cond} is shown by green dashed lines. All conductivity is in unit of $e^2/h$. (a) $\sigma_H$ and $\sigma_\text{H}^\text{T}$ as a function of $m$ varying from $-2.5$ to $2.5$, with $f({\bf k})=1$ and $\gamma=0.05$. (b) $\sigma_H$ and $\sigma_\text{H}^\text{T}$ as a function of $\gamma$ varying from $0$ to $1$, with $f({\bf k})=1$ and $m=1$ fixed. (c) $\sigma_H$ and $\sigma_\text{H}^\text{T}$ as a function of $a$ varying from $0$ to $0.5$, when $f({\bf k})$ is fixed at $1-a(\sin (ak_x)+\sin (ak_y))/2$, and with $\gamma=0.05$ and $m=1$ fixed. }\label{Hall}
\end{figure}

Let us present a more detailed analysis of Eq. \ref{Hall_cond}. First of all, when the momentum dependence of $\gamma_{{\bf k}}$ can be ignored, we have $\partial_{{\bf k}}{\bf d}_{{\bf k}}=\partial_{{\bf k}}({\bf d}_{{\bf k}}+i\boldsymbol{\gamma}_{{\bf k}})=\partial_{{\bf k}}{\bf h}_{{\bf k}}$, and therefore
\begin{equation}
\Omega_{xy}({\bf k})= \hat{{\bf h}}_{{\bf k}}\times (\partial_{k_x}\hat{{\bf h}}_{{\bf k}}\times \partial_{k_y}\hat{{\bf h}}_{{\bf k}}).\label{Omega_xy2}
\end{equation}
With Eq. \ref{Omega_xy2} one can show that 
\begin{align}
\Omega_{xy}({\bf k})&=\epsilon_{ij}\text{Tr}[\hat{P}^+_{{\bf k}}(\partial_{k_i}\hat{P}^+_{{\bf k}})(\partial_{k_j}\hat{P}^+_{{\bf k}})]\nonumber\\
&=\nabla_{{\bf k}}\times {\bf A}_{xy}({\bf k}),
\end{align}
where the Berry connection ${\bf A}_{xy}({\bf k})$ is defined as ${\bf A}_{xy}({\bf k})=-i\langle u_{+,{\bf k}}^L|\nabla_{\bf k}|u_{+,{\bf k}}^R\rangle$. It can be shown that both $\int \Omega_{xy}d^2{\bf k}$ and $\int \Omega^*_{xy}d^2{\bf k}$ are quantized \cite{Shen, Wang}. Secondly, for the $\nu_{{\bf k}}$ term, because both $\lambda^\text{R}_{{\bf k}}\pm h^0_{{\bf k}}$ are positive, $\nu_{{\bf k}}$ in general can take any value between $[0,1]$, and $\nu_{{\bf k}}\rightarrow 1$ when $|\lambda^\text{I}_{{\bf k}}\pm\gamma^0_{{\bf k}}| \rightarrow 0$. Note that in most cases, the Berry curvature is not uniformly distributed in the Brillouin zone but is concentrated in certain regime. Therefore, in order to obtain a nearly quantized Hall conductance, it requires (i) $\gamma_{{\bf k}}$ to be as smooth as possible and (ii) $|\gamma^0_{{\bf k}}\pm\lambda^\text{I}_{{\bf k}}|$ to be as small as possible in the regime where the Berry curvature is concentrated. 

\textit{Examples.} To illustrate how Hall conductivity deviates from unity when these two conditions above are violated, we take a specified model as follows
\begin{align}
{\cal H}_{\bf k}=&-i\gamma{\it I}+\sin k_x\sigma_x+(\sin k_y+i\gamma f({\bf k}))\sigma_y\nonumber\\
&+(m+\cos k_x+\cos k_y)\sigma_z\label{Eqn:Model}.
\end{align}
In the Hermitian limit $\gamma\rightarrow 0$, the system recovers the model describing the quantum anomalous Hall effect. For $0<m<1$, $\sigma_H=1$, and for $-1<m<0$, $\sigma_H=-1$. In these two regions, the model describes a topological Chern insulator. For $m>2$ and $m<-2$, $\sigma_H=0$ and the model describes a topological trivial insulator.
Here $f({\bf k})$ satisfies the constraint $|f({\bf k})|\leq 1$, under which the imaginary part of all eigen-values are negative. The Hall conductances for several different cases of this model are calculated and shown in Fig. \ref{Hall}. We display both the full numerical result for $\sigma_\text{H}$ (red solid line) and the topological contribution $\sigma_\text{H}^\text{T}$ (green dashed line) given by formalism Eq. \ref{Hall_cond}, and the difference between them stands for the bulk contribution. 

First of all, as we know from the discussion above that a constant $f({\bf k})$ results in a quantized $\int d^2{\bf k}  \Omega_{xy}({\bf k})$, therefore we first consider the $f({\bf k})=1$ case. In Fig. \ref{Hall}(a) we show the Hall conductance as a function of $m$ in the presence of a small $\gamma=0.05$. One can see that $\sigma_\text{H}^\text{T}$ still displays a nearly quantized Hall conductance when $m$ is away from the topological transition point in the Hermitian limit, i.e. $m=0,\pm 2$, and the change from $\sigma_\text{H}^\text{T}=0$ to $\sigma_\text{H}^\text{T}=\pm 1$ is a smooth crossover rather than a sharp jump. The bulk contribution is most significant when $m$ is within $\pm \gamma$ of $m=0,\pm 2$, when the system is not gapped according our definition based on the density-of-state. 

In Fig. \ref{Hall}(b) we fix $f({\bf k})=1$ and $m=1$ and show how the Hall conductance depends on $\gamma$. One can see that on one hand, the topological contribution $\sigma_\text{H}^\text{T}$ decreases as $\gamma$ increases. This is because although 
$\int d^2{\bf k}\Omega_{xy}({\bf k})$ remains quantized, $\nu_{{\bf k}}$ becomes smaller as $\gamma$ increases. On the other hand, we can also see that the bulk contribution also increases when $\gamma$ increases. In the inset of Fig. \ref{Hall}(b), we show that for small $\gamma$, both $\sigma_\text{H}$ and $\sigma^\text{T}_\text{H}$ deviate from unity linearly with $\gamma$. 

Finally, we show that the ${\bf k}$ dependence in $\gamma_{\bf k}$ can affect the quantization of $\int d^2{\bf k}\Omega_{xy}({\bf k})$, and therefore, a stronger momentum dependence of $\gamma({\bf k})$ will lead to a larger deviation of $\sigma_\text{H}^\text{T}$ from a quantized value. Here we take $f({\bf k})=1-(\sin(a k_x)+\sin(a k_y))/2$, and the increase of $a$ increases the momentum dependence of $\gamma_{{\bf k}}$. In Fig. \ref{Hall}(c) we show that $\sigma_\text{H}^\text{T}$ decreases when $a$ increases and $\gamma$ is fixed. On the other hand, because the overall strength of the non-Hermitian part does not increases and the typical energy scale of the imaginary part of eigen-energies does not increase with the increasing of $a$, the bulk contribution does not increase. 
 
\textit{Conclusions.} In summary, this work quantifies how the Hall conductance of a non-Hermitian Chern insulator deviates from quantized value even when the non-Hermitian Chern number is quantized. The Hall conductance in general contains the bulk contribution and the topological contribution. The bulk contribution is non-universal and increases with the increasing of the strength of the non-Hermitian terms, because it increases the density-of-state inside the gap. We present a non-Hermitian version of the Thouless-Kohmoto-Nightingale-de Nijs formula that shows that the topological contribution deviates from quantized value either when the strength of the non-Hermitian terms increases, because it increases the decay of edge modes, or when the momentum dependence of the non-Hermitian terms increases. As the non-Hermitian terms can be controlled in many synthetic topological systems, our results can be directly verified in experiments in near future. 

\textit{Acknowledgement.} This work is NSFC under Grant No. 11604225 (YC) and No. 11734010 (YC and HZ), MOST under Grant No. 2016YFA0301600 (HZ), Foundation of Beijing Education Committees under Grants No. KM201710028004 (YC).  We thank Zhong Wang, Huitao Shen for very helpful discussions. 



\end{document}